# Buried dislocation networks designed to organize the growth of III-V semiconductor nanostructures


José Coelho, Gilles Patriarche, Frank Glas, Guillaume Saint-Girons, Isabelle Sagnes and

Ludovic Largeau

Laboratoire de Photonique et de Nanostructures, CNRS, Route de Nozay, F-91460

Marcoussis, France



We first report a detailed transmission electron microscopy study of dislocation networks (DNs) formed at shallowly buried interfaces obtained by bonding two GaAs crystals between which we establish in a controlled manner a twist and a tilt around a <110> direction. For large enough twists, the DN consists of a two-dimensional network of screw dislocations accommodating mainly the twist and of a one-dimensional network of mixed dislocations accommodating mainly the tilt. We show that in addition the mixed dislocations accommodate part of the twist and we observe and explain slight unexpected disorientations of the screw dislocations with respect to the <110> directions. By performing a quantitative analysis of the whole DN, we propose a coherent interpretation of these observations which also provides data inaccessible by direct experiments. When the twist is small enough, one screw subnetwork vanishes. The surface strain field induced by such DNs has been used to pilot the lateral ordering of GaAs and InGaAs nanostructures during metalorganic vapor phase epitaxy. We prove that the dimensions and orientations of the nanostructures are correlated with those of the cells of the




underlying DN and explain how the interface dislocation structure governs the formation of the nanostructures.

61.72.Lk, 61.72.Mm, 68.37.Lp, 68.65.-k

# I. INTRODUCTION

Controlling the spontaneous periodic ordering of nanostructures could open the way to the realization of numerous new devices. Indeed, if ordering with adjustable periods and orientations were achieved, it might become possible to control the size, shape, density and spatial distribution of the nanostructures. The use of such a technique would for instance permit to increase the density of InGaAs quantum dots (QDs) grown by metal organic vapor phase epitaxy (MOVPE) on a GaAs substrate, in order to enhance the modal gain of their optical fundamental transition and then to obtain a laser operating at the 1.3 μm wavelength. The technique could also be applied to new devices, such as single photon sources for quantum cryptography based on isolated QDs. However, although spontaneous formation of III-V QDs has been obtained long ago,[1] the tailoring of their geometric properties has remained an elusive task. To put it briefly, the aim is to go from 'self-assembling' to 'self-organization' (controlled organization).



A promising way to order nanostructures is to use the strain field induced at the surface of a specimen by a periodic dislocation network (DN) shallowly buried and parallel to the surface, which has been predicted (via elasticity calculations) to generate preferential nucleation sites for QDs.[2][3] By choosing appropriate DN periodicities, it should be possible to order laterally QDs having identical shapes an sizes. Such a subsurface DN can be obtained by wafer bonding, a technique which has the advantage of not leading to the formation of any threading dislocation which could affect the optical properties of the subsequently grown layers.[4][5][6] Moreover, the efficient charge carrier confinement in the QDs should strongly reduce their optical sensitivity to the presence of defects, and in particular to the dislocations of a shallowly buried DN.[7] So far, lateral organization of QDs mediated by an underlying DN has been reported for metals[8] and for Germanium on Silicon[9] but not for III-V materials.

We recently reported a major step towards the control of III-V QDs grown on GaAs by using this technique, namely the lateral organization of specific GaAs and InGaAs nanostructures (to be recalled below).[10] Here we first present a detailed transmission electron microscopy (TEM) study of our DNs performed before growing the nanostructures and a quantitative interpretation of their structure. Then, using new observations of the DNs and of the ordering of the nanostructures, we demonstrate and explain their correlations with the underlying DN.



## II. FORMATION OF THE DISLOCATION NETWORKS BY CRYSTAL BONDING

DNs form in particular at the planar interface between two identical but differently oriented crystals, which is a type of grain boundary (GB). In the present work, we consider slightly disoriented III-V crystals having the same lattice parameter $a$, with interfaces close to a (001) plane. In III-V materials, the dislocations have Burgers vectors **b** of the $a/2<110>$ type. The two disorientations which may happen are a twist (*i.e.* a rotation around an axis orthogonal to the interface) and a tilt (*i.e.* a rotation around an axis lying in the interface); the latter occurs when at least one of the two bonded surfaces is vicinal. Taken independently, the twist may be accommodated by a square two-dimensional (2D) network of screw dislocations. On the other hand, tilt accommodation requires Burgers vector with components normal to the GB[11]. From results previously obtained for Si/Si[12] and GaAs/InP bonding[6], we may expect these to be provided by a one-dimensional (1D) network of mixed dislocations oriented along the tilt axis (the line orthogonal to the maximum slope of the interface). For both kinds of ideal GBs, the dislocation periodicity is:

$$D = \tfrac{1}{2}\, |b'| / \sin(\theta/2) \,, \quad (1)$$

where $\theta$ is the disorientation angle and $|b'|$ the modulus of the component of the Burgers vector allowing the accommodation of the crystalline discontinuity ($|b'| = a\sqrt{2}/2$ for screw dislocations and the component $|b'| = a/2$ normal to the GB for the mixed dislocations; see Table I).



In order to obtain DNs shallowly buried and parallel to the surface, we use the epitaxial wafer bonding technique.[4][5][6] The first crystal to be bonded is a standard GaAs 'host' substrate. The second crystal is a GaAs substrate on which we first grow an AlGaAs layer to be used as an etch-stop layer during subsequent chemical selective etching, followed by a 20 nm thick GaAs layer to be transposed on the host substrate. After cleaning and deoxidizing, the mirror-like surfaces of the two samples are put in contact under mechanical pressure (between 10 and 100 kg/cm²), at room temperature and with controlled disorientations. The tilt is established by using commercial wafers having a vicinality controlled to within ± 0.1 degree. To obtain a twist between the two bonded substrates, we first cut with a saw square pieces of a wafer to obtain sides having the desired disorientation with respect to the <110> cleavage directions. We then put in contact a sawn square and square simply cleaved along the <110> directions to which it will be bonded. We align their sides by propping them against a wedge so that the desired twist is imposed. This method allows twist control to within ± 0.1 degree. The two samples are then annealed at 600°C during one hour under a flow of nitrogen. At this temperature, covalent bonds form between the two crystals at the interface and the DNs accommodating the disorientations appear[6] (since no change of the interface structure is observed when the annealing time is varied between 30 minutes and two hours, we must have reached the equilibrium configuration). The structure is then thinned from the side of the substrate containing the etch-stop layer down to the latter, which is subsequently removed to leave a final assembly composed of the thin GaAs layer bonded to the host substrate.



# III. DETAILED STUDY OF THE INTERFACIAL DISLOCATION NETWORKS

## A. Basic structure of the networks

In this Section, we study in detail the DNs obtained by bonding two substrates with vicinal surfaces identically disoriented around an in-plane <100> direction. Before bonding, we rotate them by 90 degrees so that their 1D networks of surface steps are nearly orthogonal; hence, the resultant tilt axis is close to a <110> direction. Moreover, we establish between them a small additional twist (a few degrees at most). To make observations easier, the study is carried out before the growth of the nanostructures (see Sec. IV), which does not modify the DNs. We use mainly plan-view TEM images of thin specimens containing overlapping portions of the bonded crystals; this is the choice technique for imaging large areas of complex dislocation networks, since it allows the interface to occupy the whole image field (whereas it projects as a line in cross-sectional images).

We might expect the interfacial DNs to be a simple superposition of the two ideal networks of screw and mixed dislocations described above. Indeed, as a first approximation, we find that the interfacial GBs of our composite substrates



systematically contain a 1D network of mixed dislocations and what appears to be a 2D network of screw dislocations (for the moment, we retain this denomination; deviations from pure screw character will be discussed below). This is demonstrated in Fig. 1, which shows two TEM dark-field weak-beam plane-view images of the GB, taken with two perpendicular 220-type diffraction vectors **g** allowing to image either the 1D network of mixed dislocations and one of the components of the 2D network of screw dislocations [Fig. 1(a)], or the second component of the screw network [Fig. 1(b)]. These DNs are schematized in Fig. 2. Note that in Fig.1 (a), the screw dislocations are more contrasted than the mixed ones (probably due to their larger $|\mathbf{g}.\mathbf{b}|$ value) and that in Fig. 1(b), each crossing of a mixed dislocation shifts the screw dislocations by half a period. This interaction makes the mixed dislocations adopt a configuration consituted by a sequence of segments disoriented from the average line direction (Fig. 2). Such energy-minimizing shifts and reorientations have already been reported for interactions between 60 degree dislocations and screw or edge networks [6 12 13]. On the other hand, we do not observe dislocation dissociation. This is not surprising since in GaAs the typical dissociation length is only about 0.5 nm. Anyway, dissociation on such small scales, well below the various charcteristic lengths involved (dislocation periodicities, layer thicknesses) would affect neither the quantitative analysis carried out in Sections III and IV nor the growth of the nanostructures (Section IV).

In this sample, all dislocations are roughly oriented along <110> directions and the periodicities of the two DNs are close. This orientation is of course expected for screw



dislocations, but also for the mixed ones, given the vicinalities of the bonded surfaces (here and in what follows, the 'orientation' of a mixed dislocation must be understood as its average orientation, unless its various segments are specifically considered). Since we do not observe any influence of the polarities of the substrates on our results, we need not distinguishing between the absolute [110] and $[\bar{1}10]$ directions of either substrate and we arbitrarily label the orientations of the dislocations of Fig. 1(a) and 1(b) as [110] and $[\bar{1}10]$, respectively; quantities pertaining to the two screw subnetworks will be labeled with indices 1 and 2, respectively. To prevent any confusion between crystal disorientations and DN disorientations, the former are noted with Greek letters $\theta$ and the latter with $\omega$, with indices specifying the DN. Although the screw and mixed dislocations are not exactly oriented along <110> directions, they remain very close to the latter (see Sec. III.C and III.D) and it is worth keeping in mind the magnitudes of the various components of the Burgers vectors for perfect screw dislocations and for [110]-oriented mixed dislocations ('60 degree dislocations') (Table I).

The disorientations and periods measured for the various DNs are given in Table II. Note that in this Table, the figures after the ± sign are not actually error bars but mainly reflect the local fluctuations of the measured quantities. These dispersions can be accurately measured in the large thin areas of our plan-view images.



## B. A rectangular network of screw dislocations

A major deviation from the simple picture outlined in Sec. II appears readily: the two subnetworks of screw dislocations have different periodicities D1 and D2 (Table II), so that the screw DN is actually rectangular (slight deviations from orthogonality will be explored in Sec. III.C). We propose that the difference of periodicity happens because the mixed dislocations accommodate not only the tilt but also part of the twist thanks to their screw components in the GB plane. Reasoning as a first approximation on isolated and straight dislocations (dislocation interaction will be considered later), such a component must exist for any mixed dislocation because the in-plane component of its Burgers vectors is along a <100>-type direction, whereas its line is close to [110] (from now on, 'in-plane' means 'in the GB plane'). Moreover, these screw components are along this same line, and thus nearly parallel to those of the [110] screw dislocations. If they add up, accommodating a given twist requires less screw dislocations oriented along [110] (but not along $[\bar{1}10]$) and the periodicity D1 (but not D2) becomes larger than expected.

To check this hypothesis, we first measured by electron diffraction the actual twist angle as $\theta_{\text{twist}} = 1.65 \pm 0.25$ degree. From Eq. (1), the corresponding ideal *square* network of screw dislocations would have a periodicity $D_{\text{twist}} = 13.9 \pm 2.1$ nm. Table II shows indeed that $D_1 > D_{\text{twist}}$ whereas $D_2 = D_{\text{twist}}$ within experimental uncertainties. Our explanation can be confirmed quantitatively by checking that the total twists accommodated by screw Burgers vector components lying along [110] and $[\bar{1}10]$ are equal. Hence, we must have:



$$\theta_{t2} = \theta_{t1} + \theta_{tm} = \theta_{twist}, (2)$$

where angles $\theta_{t1}$, $\theta_{t2}$ and $\theta_{tm}$ are the twists accommodated respectively by the two subnetworks of the actual screw DN and by the in-plane screw components of the Burgers vector of the mixed dislocations. From Eq. (1) we get $\left| \theta_{ti} \right| = 2 \arcsin \left( a / \sqrt{2} D_i \right)$ and, using the periods given in Table II, we obtain $\theta_{t2} = 1.53 \pm 0.04$ degree, $\theta_{t1} = 1.09 \pm 0.04$ degree. The mixed dislocations have eight possible different Burgers vectors, and the twist which they collectively accommodate is $\left| \theta_{tm} \right| = 2 \arcsin \left( \left| b_{ms} \right| / 2 D_m \right)$, where $\mathbf{b}_{ms}$ is the average projection of all these vectors along the mixed dislocation line and $D_m$ the network period (Table II). Assuming that all their normal components have a given orientation (arbitrarily noted [001]), which simply means that the density of mixed dislocations is the minimum necessary to accommodate the tilt, the Burgers vectors may be either along [101], [011], [0$\bar{1}$1] or [$\bar{1}$01]. If we further assume that they are either all along [101] or [011], or all along [0$\bar{1}$1] or [$\bar{1}$01], and in each case in equal numbers, they all correspond to a twist of the same sign because they have the same projection along [110]. Then, since the mixed line is on average very close to [110], $\left| b_{ms} \right| \approx a \sqrt{2} / 4$ (Table I) and $\theta_{tm} = 0.41 \pm 0.05$ degree, so that Eq. (2) is verified.

Although our hypothesis on the selection of the Burgers vectors is thus largely justified, experimental uncertainties leave open the possibility that a small fraction of their screw components might cancel each other. This will be discussed in detail in Sec. III.D.



Moreover, dislocation interaction leads us to modify slightly the previous reasoning, performed as if each mixed dislocation had a well-defined Burgers vector. It can however be seen easily that the half period shift of the screw dislocations parallel to $[\overline{1}10]$ upon crossing a mixed dislocation (Fig. 1(b)) makes the Burgers vector of each segment of the latter take two values. Only two such pairs exist: either $a/2[101]$ and $a/2[011]$, or $a/2[0\overline{1}1]$ and $a/2[\overline{1}01]$ (Fig. 3). Since only the net Burgers vector matters for the accommodation of crystal discontinuities and since, in each pair, the screw component is the same, our previous demonstration remains valid; we only need to specify that the second pair is much less frequent than the first. Note also that dislocation interaction constrains all the Burgers vectors of screw subnetwork #2 to be the same.

We thus conclude that the mixed dislocations accommodate part of the twist via their screw components in the GB plane and that this induces a difference of periodicity for the two screw subnetworks. This phenomenon is however not observed when $D_l$ is very small because the mixed dislocations are then insufficiently numerous to accommodate a significant portion of the twist.

## C. Disorientation of the dislocations with respect to the <110> directions



The rectangular geometry of the screw DN discussed in Sec. III.B is the most obvious deviation from the ideal picture presented in Sec III.A. Besides this, we observe subtler discrepancies. In particular, each subnetwork of screw dislocations is slightly disoriented with respect to its near-<110> direction (Table II). Hence, these dislocations are not perfect screw dislocations. These disorientations imply the existence of a small Burgers vector edge component $\mathbf{b}_{ei}$ in the GB plane for both subnetworks $i$ =1,2. Since these nearly orthogonal edge components cannot mutually cancel, there are two possibilities. They might accommodate a slight lattice mismatch between the bonded crystals. If not, they must be cancelled by another edge component of the full interface DN. We shall examine these two hypotheses in turn.

The lattice mismatch which would be accommodated by the edge components of screw subnetwork #$i$, disorientated by $\omega_i$, is:

$$f_i = b_{ei}/D_i = \left(a \sin \omega_i\right)\Big/\left(\sqrt{2}D_i\right). \ (3)$$

Since each screw subnetwork would only accommodate a mismatch normal to its quasi-<110> line direction, we should have $\sin \omega_1 / D_1 = \sin \omega_2 / D_2$, which is not excluded by our data (Table II). For example, the [110]-oriented subnetwork would accommodate misfit $f_1 = (9.3 \pm 2.6) \times 10^{-4}$ along $[\bar{1}10]$.

The only possible cause of misfit between the two bonded GaAs substrates seems to be a doping difference. However, our two substrates have identical nominal doping levels.



This is however not critical, since doping-induced misfits are very small compared with the abovementioned figure: for instance, using the results of Bassignana et al.,[14] we find that the misfit expected between our substrates and undoped GaAs is less than $4 \times 10^{-5}$. Hence doping, and more generally lattice misfit, cannot explain the disorientation of the screw subnetworks.

We thus turn to our second hypothesis: the edge components of the screw dislocations cancel (or are cancelled by) other edge components. The only possible source of the latter is the network of mixed dislocations. Since, as already noticed, the in-plane components of the Burgers vectors of each segment of the latter are approximately at 45 degrees of the average mixed line direction (and anyway never parallel to the segment) (Fig. 3), these components indeed exist (Table I). The net edge component of the mixed dislocations depends on the relative magnitudes of the projections of the sum of their Burgers vectors along and perpendicular to their line direction. Unfortunately, their poor contrast and their waviness hinder any precise measurement of their average orientation (Fig. 1(a)). We may nevertheless assert that the latter deviates from [110] by a few degrees at most. In Sec. III.D, we demonstrate that our experimental data confirm this hypothesis and allow a precise determination of the parameters involved.



## D. Global analysis of the interface dislocation networks

The orientations of the DNs are summarized in Fig. 4. The arbitrarily chosen orientation of the dislocations lines is indicated by arrows. We call $\mathbf{u}_{hkl}$ the unit vector along direction $[hkl]$. All directions refer to one (arbitrarily chosen) of the two crystals. We call $\omega_1$ and $\omega_2$ the disorientations of each screw line with respect to its neighboring <110> direction and $\omega_m$ the average disorientation of the mixed line with respect to [110]. Our data show that all these angles are small and that $\omega_1$ and $\omega_2$ are of the same sign (Table II). Since the twist is mainly accommodated by the screw dislocations, the projections of the Burgers vectors of the latter along their lines must be of the same sign; we arbitrarily choose these vectors to be $\mathbf{b}_1 = \left( a/\sqrt{2} \right) \mathbf{u}_{110}$ and $\mathbf{b}_2 = \left( a/\sqrt{2} \right) \mathbf{u}_{\bar{1}10}$, respectively (Fig. 4).

A full description of the interface would include, in addition to the measured quantities (periods of the screw and mixed networks, disorientation of the screw lines, twist angle), the determination of the tilt angle $\theta_{tilt}$, of the disorientation $\omega_m$ of the mixed dislocations and of the relative probabilities of their eight possible Burgers vectors $\mathbf{b}_m$. To solve this problem, we make only one assumption: as already discussed above, we suppose that the <001> normal components of the Burgers vectors of the various mixed dislocations all have the same sign which, without loss of generality, we shall take as positive along [001]. This assumption is much less restrictive than that adopted in Sec. III.B, since we



now allow four possible Burgers vectors, namely $\left(a/\sqrt{2}\,\right)\mathbf{u}_{101}$, $\left(a/\sqrt{2}\,\right)\mathbf{u}_{011}$,

$\left(a/\sqrt{2}\,\right)\mathbf{u}_{\bar{1}01}$ and $\left(a/\sqrt{2}\,\right)\mathbf{u}_{0\bar{1}1}$ and make no assumption about their relative

probabilities, which we note $p_{10}$, $p_{01}$, $p_{\bar{1}0}$ and $p_{0\bar{1}}$, respectively. These probabilities

verify:

$$0 \le p_{ij} \le 1 \quad \text{and} \quad p_{10} + p_{01} + p_{\bar{1}0} + p_{0\bar{1}} = 1 \,. \quad (4)$$

Because of interaction with screw network #2, the Burgers vectors do not remain uniform

along each mixed dislocation but alternate (see Sec. III.B and Fig. 3), so that the

probabilities must be understood as averages both along and between dislocations.

The geometry of all combinations of periodic linear interfacial DNs compatible with a

given disorientation and misfit between two crystals is prescribed by Frank's formula[11],

which states that if a given vector $\mathbf{V}$ of one crystal transforms in the other crystal into

vector $\mathbf{V}'$ (having the same crystal coordinates), the 'closure defect' $\mathbf{B(V)} = \mathbf{V} - \mathbf{V}'$

verifies:

$$\mathbf{B(V)} = \sum_{p} c_{p}\,(\mathbf{V})\,\mathbf{b}_{p} \,. \quad (5)$$

In Eq. (5), the sum extends to all DNs $p$ whose dislocations have Burgers vectors $\mathbf{b}_{p}$,

and $c_{p}\,(\mathbf{V})$ is the number of such dislocations intersecting $\mathbf{V}$; calling $\mathbf{n}$ the interface

normal pointing from the crystal containing $\mathbf{V}$ to that containing $\mathbf{V}'$, $c_{p}\,(\mathbf{V})$ is counted

positive if $\mathbf{V} \times \mathbf{n}$ has a positive component along the oriented dislocation line. We take $\mathbf{n}$

as having a (large) positive component along $\mathbf{u}_{001}$. Applying Frank's formula to the two



non-colinear unit vectors $\mathbf{u}_{110}$ and $\mathbf{u}_{\bar{1}10}$ and to the three DNs (screw #1 and #2 and mixed), and ignoring for the moment that, due to interaction, the mixed dislocations are not straight, we readily get:

$$\mathbf{B}(\mathbf{u}_{110}) = -\frac{\sin\omega_1}{D_1}\frac{a}{\sqrt{2}}\mathbf{u}_{110} - \frac{\cos\omega_2}{D_2}\frac{a}{\sqrt{2}}\mathbf{u}_{\bar{1}10} - \frac{\sin\omega_m}{D_m}\left[\frac{a\sqrt{2}}{4}\left(\alpha\mathbf{u}_{110} + \beta\mathbf{u}_{\bar{1}10}\right) + \frac{a}{2}\mathbf{n}\right], (6)$$

$$\mathbf{B}(\mathbf{u}_{\bar{1}10}) = \frac{\cos\omega_1}{D_1}\frac{a}{\sqrt{2}}\mathbf{u}_{110} - \frac{\sin\omega_2}{D_2}\frac{a}{\sqrt{2}}\mathbf{u}_{\bar{1}10} + \frac{\cos\omega_m}{D_m}\left[\frac{a\sqrt{2}}{4}\left(\alpha\mathbf{u}_{110} + \beta\mathbf{u}_{\bar{1}10}\right) + \frac{a}{2}\mathbf{n}\right], (7)$$

where:

$$\alpha = p_{10} - p_{\bar{1}0} + p_{01} - p_{0\bar{1}}, (8)$$

$$\beta = p_{\bar{1}0} - p_{10} + p_{01} - p_{0\bar{1}}. (9)$$

Coefficients $\alpha$ and $\beta$ characterize the projection in the interface plane of the average Burgers vector $\overline{\mathbf{b}}_m$ of the mixed dislocations:

$$\overline{\mathbf{b}}_m = \frac{a\sqrt{2}}{4}\left(\alpha\mathbf{u}_{110} + \beta\mathbf{u}_{\bar{1}10}\right) + \frac{a}{2}\mathbf{u}_{001}. (10)$$

Since angles $\omega_1, \omega_2, \omega_m, \theta_{twist}$ and $\theta_{tilt}$, are all small, we shall retain only the terms of first order in angles. Although not necessary, this simplifies considerably the discussion. If $\theta_{twist} > 0$ corresponds to a rotation from $\mathbf{u}_{110}$ towards $\mathbf{u}_{\bar{1}10}$, the closure defects then become:

$$\mathbf{B}(\mathbf{u}_{110}) = -\theta_{twist}\,\mathbf{u}_{\bar{1}10} + \omega_m\,\theta_{tilt}\,\mathbf{u}_{001}, (11)$$

$$\mathbf{B}(\mathbf{u}_{\bar{1}10}) = \theta_{twist}\,\mathbf{u}_{110} - \theta_{tilt}\,\mathbf{u}_{001}. (12)$$

Projecting Eqs. (6), (7) on $\mathbf{u}_{001}$, $\mathbf{u}_{110}$ and $\mathbf{u}_{\bar{1}10}$ and setting $d_0 = a/\sqrt{2}$ yields:



$$\theta_{tilt} = -a/2 D_m \, , \, (13)$$

$$\alpha \, \omega_m = -2 \, \omega_1 \, D_m \, / \, D_1 \, , \, (14)$$

$$\beta \, \omega_m = -2 D_m \, (1/D_2 - \theta_{twist}/d_0) \, , \, (15)$$

$$\alpha = -2 D_m \, (1/D_1 - \theta_{twist}/d_0 ) \, , \, (16)$$

$$\beta = 2 \, \omega_2 \, D_m \, / \, D_2 \, . \, (17)$$

Eq. (13), obtained twice by projecting along $\mathbf{u}_{001}$, is simply Eq. (1) for small tilt. That $D_m$ depends only on the tilt simply means that the screw dislocations cannot accommodate any part of the latter. Numerically, $\theta_{tilt} = (-1.02 \pm 0.10) \times 10^{-2}$, or $\theta_{tilt} = -0.58 \pm 0.06$ degree (in agreement with a value of $-0.71 \pm 0.14$ degree calculated from substrate vicinalities of $0.5 \pm 0.1$ degree ).

From the four remaining equations we want to extract the three unknown $\omega_m$, $\alpha$ and $\beta$. From a practical point of view, we note that the DN periods are measured much more precisely than the disorientations (Table II). We thus found that, somewhat surprisingly, the best procedure is to first obtain from Eqs. (14)-(17) a refined value of $\theta_{twist}$ (treated as a fourth unknown) rather than using its measured value. This calculated value depends only weakly on the measured angles and can then be used in the equations to obtain accurate values of the other parameters. From (14)-(17), we get:

$$\omega_m = \frac{\omega_1}{1 - \theta_{twist} \, D_1/d_0} = -\frac{1 - \theta_{twist} \, D_2/d_0}{\omega_2} \, . \, (18)$$

Hence, $\theta_{twist}$ verifies:



$$D_1 D_2\, \theta_{twist}^2 - d_0\,(D_1 + D_2)\,\theta_{twist} + (1 + \omega_1\,\omega_2)\,d_0^2 = 0 \; . \; (19)$$

The two possible solutions of (19) are:

$$\theta_{twist}^- = d_0\left(\frac{1}{D_2} - \frac{\omega_1\omega_2}{D_1 - D_2}\right) \quad \text{and} \quad \theta_{twist}^+ = d_0\left(\frac{1}{D_1} + \frac{\omega_1\omega_2}{D_1 - D_2}\right). \; (20)$$

These values depend little on angles $\omega_1$ and $\omega_2$ because only via the product of these angles and because $D_1$ and $D_2$ are noticeably different. Indeed, the data of Table II yield the following narrow ranges: $\theta_{twist}^- = (2.65 \pm 0.07) \times 10^{-2}$ and $\theta_{twist}^+ = (1.92 \pm 0.07) \times 10^{-2}$. Whereas the '+' range is incompatible with our experimental determination, the '-' range lies fully within our experimental values, which strengthens our argument. Hence $\theta_{twist} = \theta_{twist}^-$ and our refined value is $\theta_{twist} = (2.65 \pm 0.07) \times 10^{-2}$, or $\theta_{twist} = 1.52 \pm 0.04$ degree.

Inserting this value and the experimental values of $D_1$ and $D_m$ in (16) leads to $0.78 \leq \alpha \leq 1.34$. However, (8), (9) and (4) indicate that $-1 \leq \alpha \leq 1$ and $-1 \leq \beta \leq 1$. Finally: $0.78 \leq \alpha \leq 1$. From (17) and the experimental values of $D_2$, $D_m$ and $\omega_2$, we get: $0.023 \leq \beta \leq 0.277$.

Noting that, from (8), (9) and (4):

$$p_{10} + p_{01} = (1 + \alpha)/2 \; , \; (21a)$$

$$p_{\bar{1}0} + p_{01} = (1 + \beta)/2 \; , \; (21b)$$

we finally obtain the following set of constraints on the Burgers vector probabilities:



$$\begin{cases} 0.89 \leq p_{10} + p_{01} \leq 1 \\ p_{01} \geq 0.4 \\ p_{10} \leq 0.6 \\ 0 \leq p_{\bar{1}0} + p_{0\bar{1}} \leq 0.11 \end{cases} \qquad .$$

This specifies the deviations allowed by the uncertainties on our experimental data from

the simple hypotheses adopted in Sec. III.B, which correspond to $p_{10} = p_{01} = 0.5$,

$p_{\bar{1}0} = p_{0\bar{1}} = 0$. Moreover, (18) now provides the average disorientation of the mixed

network. We find $\omega_m = (-13.6 \pm 6.2) \times 10^{-2}$, or $\omega_m = -7.9 \pm 3.5$ degree. This implies

that $\omega_m$ is necessary non-zero and has a sign of opposite to that of $\omega_1$ and $\omega_2$. Detailed

examination of the geometry of the DN shows that these results are modified neither if

the broken line character of the mixed dislocations (Fig. 3) is taken into account nor if

other possible signs of BV comonents are considered.

To summarize, our detailed *ab initio* quantitative analysis confirms and refines the

conclusions of the simplified analysis carried out in Sec. III.B and III.C. A 2D network of

'quasi-screw' dislocations accommodates mainly the twist. However, this network is

neither square nor perfectly oriented along the <110> directions. The tilt is

accommodated by a 1D network of mixed dislocations, whose individual Burgers vectors

may *a priori* have four possible components in the [001] plane. Actually, the average

Burgers vector has a large component $\alpha$ along [110], so that the mixed dislocations also

have a large screw component and accommodate a significant part of the twist, thereby

affording an increase of the period of the subnetwork of screw dislocations oriented close

to [110]. The driving force for this Burgers vector selection must be the reduced interface



energy accompanying this increased periodicity. The disorientation of the mixed

dislocations with respect to [110] and the component $\beta$ of their Burgers vector along

$[\bar{1}10]$ generate an edge component. This component is cancelled by the small edge

components generated by the slight disorientations of the screw dislocations with respect

to their respective neighboring <110> directions.

## IV. ORGANIZATION OF III-V NANOSTRUCTURES VIA THE DISLOCATION NETWORK

### A. Formation of the nanostructures

In order to study the influence of the subsurface DN, we grew by MOVPE on such

composite twist-tilt bonded substrates a sequence of layers which would give rise to the

formation of QDs on standard substrates[10]. The whole structure is shown in Fig. 5:

starting from the bottom, we find the host GaAs substrate (#1) and the bonded GaAs

layer (#2). Their interface is the GB, where the large dark spots are due to the strain fields

of the dislocations and the small ones to cavities (resulting from the non-planarity of the

surfaces put in contact) or to segregated impurities. As previously observed for other

bonded III-V crystals,[6] the dislocations constitute a planar network which remains

confined to the GB and do not propagate in the surrounding layers. The grown layers are



above layer #2. No QD is observed in this sample. However, both the GaAs buffer layer (#3) and the $In_xGa_{1-x}As$ alloy layer (#4) exhibit thickness modulations, to be discussed below. Finally, a thin GaAs layer (#5) covers the entire structure. Using our previous work,[15] we determined the composition of layer #4 from the TEM 200 dark field image intensity ratio between this layer and the GaAs layers. We found an average In composition $x = 0.31 \pm 0.02$.

The presence of a dark line at the #2/#3 GaAs/GaAs interface might seem surprising. However, the top of layer #2, on which growth is started, is obtained by chemical etching and cannot have the quality of standard 'epi-ready' wafers. Secondary ion mass spectroscopy shows that impurity levels as low as $10^{18}$ cm$^{-3}$ suffice to produce such features.

From such images, it appears readily that the thickness modulations, which affect both the GaAs buffer and the InGaAs layer, are not randomly distributed: for instance, thicker InGaAs grows in the valleys of the GaAs layer. Since moreover their dimensions, modulation periods and modulation amplitudes are of the order of between 1 and 100 nm, these features truly constitute III-V nanostructures. These nanostructures are clearly the direct effect of the underlying dislocations during growth, and are not mediated by a possible undulation of the initial growth surface, namely the top of layer #2; indeed, the latter exhibits a negligible corrugation (we measure by atomic force microscopy (AFM) a typical rms roughness of only 0.3 nm). However, images such as Fig. 5 allow a detailed study neither of the organization of the nanostructures nor of their relationship with the



underlying dislocations. In the next two sections, we first study the dislocation structure of the samples used for growth. We then demonstrate that the nanostructures are spatially correlated to the DNs and we discuss the origin of this correlation.

**B. Dislocation structure**

The composite substrates used for growing the nanostructures were obtained in the same way as those studied in Sec. III. In particular, we used two substrates disoriented around <100> directions which we bonded with their surface steps networks orthogonal to each other. However, in order to obtain DNs networks with spacings of the order of typical distances between standardly grown QDs, we selected much smaller tilt and twist angles.

Figure 6(a) is a TEM $\overline{2}20$ weak-beam image of the sample of Fig. 5 which reveals chiefly the high strain field localized close to the dislocation cores and not the more diffuse strain field associated with the growth of the strained nanostructures. Although they might appear different, the dislocations visible in this micrograph have essentially the same geometry, schematized in Fig. 6(b), as those studied in Sec. III: we observe a 1D network of mixed dislocations of average orientation close to [110] and short segments of screw dislocations along $[\overline{1}10]$ (subnetwork #2 of Sec. III). These segments and the broken line character of the mixed dislocations result again from the interaction



between the two families (Fig. 3); in particular, each screw dislocation shifts by about half a period upon crossing a mixed line. Compared with Sec. III, the smaller crystal disorientations induce larger DN periods: here, $D_1 = 261 \pm 61$ nm and $D_m = 50 \pm 15$ nm, corresponding to a twist of $0.09 \pm 0.02$ degree and to a tilt of $0.25 \pm 0.08$ degree (in agreement with expected values of $0.0 \pm 0.1$ degree and $0.21 \pm 0.07$ degree, respectively). Moreover, since the twist was set as close to zero as possible, the period $D_2$ of screw subnetwork #2 is considerably larger than the period $D_m$ of the mixed DN. Thus, the mixed segments are much easier to see than in Fig. 1(b) and we observe prolate hexagons (Fig. 6(b)).

Moreover, weak beam images formed with the orthogonal 220 reflection seem to show only the same mixed dislocations but no [110]-oriented screw dislocation, as was the case before (subnetwork #1, Fig. 1(a)). Although the presence of some screw dislocations cannot be ruled out totally, this would however not be unexpected in light of the previous study. To understand why, we must consider carefully the dislocation interactions. Fig. 7 shows a unit cell of the hexagonal DN of Fig. 6, which contains one segment of screw dislocation at the center and four segments at the corners, two segments of mixed dislocations $m1$ and $m2$ and two non-equivalent nodes N1 and N2. The only arbitrary choice (apart from the orientations of the mixed and screw lines) is that of the sign of the Burgers vector common to all the screw lines (or, equivalently, of the bonded crystal to which the crystallographic directions refer); here, for sake of comparison with TEM images, this vector is called $1\bar{1}0$, *i.e.* the opposite of the choice made in Fig. 3. The



sequence of Burgers vectors of all the segments of any mixed dislocation can be entirely determined once one of them is known: for $m1$, it is either 101-N1-011 (type $a1$) or $0\overline{1}1$-N1-$\overline{1}01$ (type $b1$) and for $m2$ either 011-N2-101 (type $a2$) or $\overline{1}01$-N2-$0\overline{1}1$ (type $b2$); as in Fig. 3, sequences of types '$a$' and '$b$' are drawn respectively above and below the mixed line. On the other hand, the sequences corresponding to $m1$ and $m2$ may *a priori* be chosen independently provided they are one of the abovementioned. From Fig. 7, which also shows the screw and edge component of each possible Burgers vector, two conclusions can be drawn. First, provided consecutive segments of each mixed dislocation are of equal length, their edge components cancel. Second, a '$b$' sequence produces large screw components; the sum of two consecutive such components is along $\left[\overline{1}\,\overline{1}0\right]$ and thus corresponds to a twist in the same rotational direction as screw subnetwork #2; on the other hand, an '$a$' sequence produces smaller screw components and a twist in the opposite direction. These two types of screw components are respectively larger and smaller than the screw component of an ideal 60 degree mixed dislocation, which is itself half as efficient as a screw dislocation to relieve twist (Table I). It is now easy to understand why screw subnetwork #1 can be totally absent. For this to happen, it suffices that the '$b$' sequences dominate and that the balance between '$b$' and '$a$' sequences produces exactly the same twist as screw subnetwork #2. From the discussion above, this can happen as soon as the ratio of the densities of mixed and screw dislocations is larger than a factor between one and two (which depends on the angle between the segments). Since we performed growth on substrates where this ratio is between 3 and 4, this mechanism can certainly operate. Actually, it is the very mechanism of Sec. III; however, the 'replacement' of a portion of screw subnetwork #1



by the screw components of the mixed DN is now pushed to its limit, the total disappearance of subnetwork #1. Indeed, with respect to interface energy, it seems highly favorable to eliminate totally one half of the standard 2D screw network. The whole interface DN geometry is then determined solely by those dislocations imaged in Fig. 6(a) and consists of prolate hexagonal cells; the long dimension of these cells (along [110]) is exactly the period $D_2$ of the screw DN and the short dimension is $h = 88 \pm 32$ nm (Fig. 6(b)). These cells are however somewhat irregular because the mixed segments are not straight, because the shift of the screw dislocations varies and because of the presence of interface cavities.

These considerations are confirmed by TEM dark field images formed with diffraction vectors **g** of 200-type (Fig. 8(a)). Since the contrast in such images is due not only to the dislocations but also to the subsequently grown layers, we show a bonded sample having the same dislocation structure but without overgrowth; later, this will also allow us to identify the growth-related features (Sec. IV.C)). The '**g.b**' rule governing dislocation contrast[16] implies that all screw segments and every second segment of each mixed dislocation should be strongly contrasted for **g** along 200, whereas the same screw segments and the rest of the mixed segments are in contrast for **g** along 020 (Fig. 7); in each case, the other half of the mixed segments should show a weak contrast[16]. Putting together the results about twist accommodation and dislocation contrast, we find that, in the case of Fig. 7, the segments of mixed dislocations in contrast for **g** along 200 have Burgers vector $a/2[\overline{1}01]$ if they belong to dominant sequences '*b*' and Burgers vector



$a/2[101]$ if they belong to minority sequences '$a$'. Conversely, the segments in contrast for **g** along 020 have Burgers vector $a/2[0\bar{1}1]$ if they belong to sequences '$b$' and Burgers vector $a/2[011]$ if they belong to sequences '$a$'. Moreover, when passing from one mixed dislocation to the next one (from $m1$ to $m2$), the pattern of contrasted and faint segments shifts laterally (by about half a period of the screw network) if the sequences of the two mixed dislocations are of the same type, whereas it does not shift if the type of sequence changes. In Fig. 8(a), we observe parallel and equidistant dark lines. Each line has steps (such as $s$ and $s'$) and all the steps displace the lines in the same direction. This can now be interpreted in the following way (Fig. 8(b)). The portion of line between two steps is constituted by mixed '$b$'-type segments separated by short screw segments (here nearly vertical and barely visible). The step is due to the absence of lateral shift between the mixed segments in contrast for two consecutive mixed dislocations, and thus corresponds to the insertion of a single '$a$'-type segment between two series of '$b$'-type segments. The step frequency depends on the respective values of twist and tilt: the larger the twist, the less frequent the steps (in Fig. 8(a), a step occurs about every four '$b$' segments). Note that in Fig. 8(a), the rest of the mixed segments appear faintly contrasted so that each hexagon can be reconstituted. Conversely, images formed with an orthogonal 200-type **g** vector show stepped lines oriented close to the complementary set of mixed segments. According to this interpretation, the strongly contrasted lines visible in Fig. 8(a) should be parallel and equidistant, and their steps should be correlated so that two hexagons with non-standard contrast patterns ($h1$ and $h2$ in Fig. 8(b)) are found between steps belonging to two neighboring lines. The former point is always verified and the latter is verified in most cases, although we also find phase shifts between steps



producing arrangements of four non-standardly contrasted hexagons (e.g. around P in Fig. 8(a)). This could be due either to the interruption of the sequences of the mixed dislocations by cavities or to the absence of a screw segment.

## C. Correlations between dislocation networks and nanostructures

Since TEM cross-sectional images such as Fig. 5 only show a section through the interface, proof of the correlation between the DN and the thickness modulations can only be obtained by studying plan-view images. Figure 9 shows standard TEM dark field images taken after growth with two perpendicular 200-type diffraction vectors. Both images show dark continuous lines. Despite slightly different DN periodicities (due to slightly different twist and tilt angles) and the contrast disturbances caused by the grown nanostructures, these lines are easily recognized as the stepped sequences of mixed and screw segments discussed in Sec. IV.B (here, a step occurs about every second mixed segment). The hexagonal cells of Figs. 6 and 7 may be reconstituted by observing the patterns of strongly and weakly contrasted segments in each image; some cells are delineated by white lines in Fig. 9(b).

In addition, Fig. 9 shows a well-defined pattern of bright areas separated by darker valleys. The three crucial points are the following:



(a) this pattern is similar in images taken with orthogonal reflections (Fig. 9 (a,b));

(b) the bright areas have the same periodicity as the underlying DN and their orientation is close to that of the cells of the DN (once allowance has been made for the irregularities of both structures);

(c) such patterns are not observed before growth: each hexagonal cell in Fig. 8(a) displays a uniform gray level (the broad variations of contrast are due to changes in diffraction conditions induced by the bending of the thin TEM specimen).

Point (a) demonstrates that the contrast is not a strain contrast but a map of the local variations of structure factor (integrated through the thickness of the specimen). These variations cannot be due to strong variations of the alloy composition, which we did not observe in the cross-sectional images (Fig. 5). Instead, they must result from the thickness variations of the GaAs and InGaAs layers observed in these same images. Indeed, since the 200 reflections have a long extinction distance whereas the specimens are relatively thin, and since moreover $In_xGa_{1-x}As$ alloys with $x \sim 0.3$ have about the same 200 structure factor as GaAs, a brighter region corresponds to a larger local thickness of GaAs or InGaAs.[16] However, since the thickness of the GaAs buffer is several times larger than the alloy thickness, the contrast pattern is dominated by the GaAs contribution. Finally, point (c) demonstrates that the correlation between DNs and thickness modulations (point (b)) was induced during growth by the former.

TEM thus demonstrates that we have managed to induce nanostructures via the strain field of the underlying DNs. These nanostructures consist of thickness modulations of the



InGaAs layer superimposed on thickness modulations of the epitaxial GaAs layer, which have the same geometry and dimensions as the cells of the underlying DNs. Yet another proof is given by AFM images (Fig. 10), which confirm the presence of the nanostructure and show that the corresponding surface height modulation is approximately 1.5 nm. Moreover, the lateral dimensions of the modulation as measured by height profiles taken along the <110> directions are identical to those observed in the TEM images and to those of the DN cells. This consistent set of observations amply proves that the strain field of the buried DN is at the origin of the ordering of the nanostructures.

## D. Interpretation

To understand how the nanostructures are induced by the DN, we must return to the dislocation structure. We observed previously that the edge component of two consecutive segments of each mixed dislocation have opposite directions. This implies that the strains induced on a given side of the interface by such a pair also alternate between dilatation and contraction (normal to the line); this is indicated by '+' and '-' in Fig. 7. This remains arbitrary as long as we have not specified relative to which crystal Fig. 7 is drawn. Let us assume that '+' corresponds to an expansion in the 'upper' crystal (which contains the grown layers).



In Fig. 11, we have drawn several hexagons of the DN in an area free of step and indicated the sign of the dilatational strain induced by each mixed segment. In this respect, screw segments remain neutral, since they only induce shear. For reason of symmetry, the dilatation must cancel at the centers of the hexagons. When the GaAs buffer is deposited, it will grow preferentially in areas where the surface lattice parameter is close to its own, namely at the central region of each hexagon (Fig. 11). This explains why the thickness of this GaAs layer is not uniform (Fig. 5), and moreover why the thick parts should coincide with these areas, as demonstrated in Sec. IV.C. According to the same mechanism, we also expect the alloy layer to grow preferentially where the surface is dilated (Fig. 11). This explains why the InGaAs alloy tends to grow in some valleys of the GaAs layer (Fig. 5). The elongated aspect of some features observed in Fig. 10 might be an indication of the chains of GaAs and InGaAs regions which appear in Fig. 11 and result from the growth of the alloy in only half the GaAs valleys.

We have not yet been attempted to measure the strain modulation at the surface (retrieving this information from TEM images is hampered by the surface relaxation of the thinned TEM specimens). Whereas this would certainly be intersting, we believe that it is not of fundamental interest in the present context, since the demonstration of a strain modulation is sufficient to ascertain the existence of a modulation of the chemical potential for the various species diffusing on the surface during growth, which is the basic ingredient in the formation of the nanostructures.



## V. SUMMARY AND CONCLUSIONS

In the first part of this work, we studied shallowly buried GaAs/GaAs interfaces obtained by wafer bonding. The interface DNs are composed of a 1D network of mixed dislocations which accommodates the tilt between the two bonded samples and of a 2D network of quasi-screw dislocations which accommodates mainly the twist. Their detailed study revealed several unexpected phenomena. First, when the mixed dislocations are only slightly disoriented with respect to a <110> direction and when the periodicity of the screw DN is not much lower than that of the mixed dislocations, the latter contribute to the accommodation of a significant part of the twist. In addition, slight disorientations of the screw dislocations with respect to the <110> directions cancel the residual in-plane edge components of the mixed dislocations due to a combination of their disorientation with respect to <110> and the value of their average Burgers vector.

We then used the strain field of such shallowly buried dislocation networks to modulate the surface potential of various atomic species during the deposition by MOVPE of a sequence of GaAs and InGaAs layers. We managed to order laterally III-V nanostructures which consist of modulations of the thicknesses of these layers. These nanostructures have lateral dimensions and orientations identical to those of the cells of the underlying DN. This result is very promising in the perspective of ordering QDs for



applications to optical emitters. We are currently trying to define conditions which will allow the organization of proper individual QDs thanks to the same type of underlying DNs.

Due to dislocation interaction, the DNs studied in this work have a more complex structure than the simple DNs proposed as potential 'growth organizers' in Refs. [2] and [3]. With respect to edge networks, they have the advantage of a geometry easily adjustable by simply changing the disorientations between crystals (and not their difference of lattice parameters). Moreover, it is precisely the interaction between dislocations which creates the alternating pattern of lattice expansion and contraction which itself leads to the formation of the nanostructures.

**ACKNOWLEDGMENTS**


We thank C. Mériadec for her expert assistance during the epitaxial bonding experiments and C. David for the AFM image. This work was supported by Région Ile de France, by SESAME project No 1377 and by Conseil Général de l'Essonne.




**TABLE I. Moduli of the Burgers vector components of ideal screw and 60 degree mixed dislocations.**

| Dislocation network | Screw component | Edge component in GB | Edge component normal to GB |
|---|---|---|---|
| screw | $a\sqrt{2}/2$ | 0 | 0 |
| 60 degree mixed | $a\sqrt{2}/4$ | $a\sqrt{2}/4$ | $a/2$ |

**TABLE II. Characteristics of the dislocations networks imaged in Fig. 1.**

| Dislocation network | Index $j$ | Nearest <110> direction | Disorientation $\omega_j$ (degree) | Period $D_j$ (nm) |
|---|---|---|---|---|
| screw #1 | 1 | [110] | $2.8 \pm 0.8$ | $21.0 \pm 0.6$ |
| screw #2 | 2 | $[\bar{1}10]$ | $2.1 \pm 1.7$ | $15.0 \pm 0.3$ |
| mixed | m | [110] | | $27.9 \pm 2.8$ |



**FIG. 1. TEM dark-field plan-view images taken in (g-4g) weak beam condition of a sample before growth of nanostructures: (a)** $g = 220$ **; (b)** $g = \overline{2}20$ **. The full and dashed arrows point respectively to screw and mixed dislocations.**

**FIG. 2. Schematics of the DNs of Fig. 1. Thick full lines: screw dislocations; dotted lines: mixed dislocations. A fine line indicates the average orientation of the latter.**

**FIG. 3. Schematics of the interaction between the mixed dislocations and screw subnetwork #2. Arrows indicate line orientation. Burgers vectors are indicated for each segment with factor a/2 omitted; for the mixed dislocation, the two possible pairs appear respectively above and below the line. Dashed line: average direction of the mixed dislocation.**

**FIG. 4. Orientations of the interfacial DNs. All directions are relative to a given crystal. Dashed lines: crystallographic directions. Full and dotted lines: dislocations. Here,** $\omega_m < 0$ **.**



**FIG 5.** TEM 002 dark-field cross-sectional image of a sample containing a buried DN, after MOVPE growth. The different layers are detailed in the text. Oval and rectangle indicate respectively an interface dislocation and an interface cavity. Note the different horizontal and vertical scales chosen to enhance the undulations of layers #3 and 4.

**FIG 6.** (a) TEM dark-field plan-view image in $\bar{2}20$ (g-4g) weak beam condition of the sample shown in Fig. 5. Two interface cavities are marked by rectangles. (b) Schematics of (a) with mixed (dotted lines) and screw (full lines) dislocations.

**FIG. 7.** Unit cell of the networks of screw and mixed (m1, m2) dislocations of samples used for growth. The inset gives the crystalline directions. Large arrows as in Fig. 3. Near each segment are indicated the possible Burgers vectors (with factor a/2 omitted) and the decomposition of their in-plane projection into screw and edge components (vectors with small arrows). Full (resp. dashed) ellipses indicate segments strongly contrasted in dark field images formed with diffraction vectors along 200 (resp. 020). '+' and '-' conventionally indicate the type of strain in the region above the interface.



**FIG 8. (a)** TEM plan-view 200 dark field image of the bonded interface in a sample without grown layers. Several hexagonal cells are highlighted and steps s, s' and non-standard set of hexagons around P are indicated. **(b)** Schematics of segments of screw and mixed dislocations contrasted strongly (full lines) and weakly (dashed lines) for diffraction vector along 200. Directions and sequence types as in Fig. 7. Dotted line: average orientation of the lines visible in (a). Scales are different in (a) and (b).

**FIG. 9.** TEM 200 dark field plan-view images of the sample shown in Figs. 5 and 6: **(a)** $g = 020$ ; **(b)** $g = 200$.

**FIG. 10.** AFM image showing the surface corrugation induced by the organized nanostructures.

**FIG. 11.** Schematics of the hexagonal cells of the DN. '+' and '-' indicate dilatation and contraction above the interface. Dashed (resp. full) ellipses indicate regions where GaAs (resp. InGaAs) preferentially grows.



Figure 1, Coelho, Phys. Rev. B.

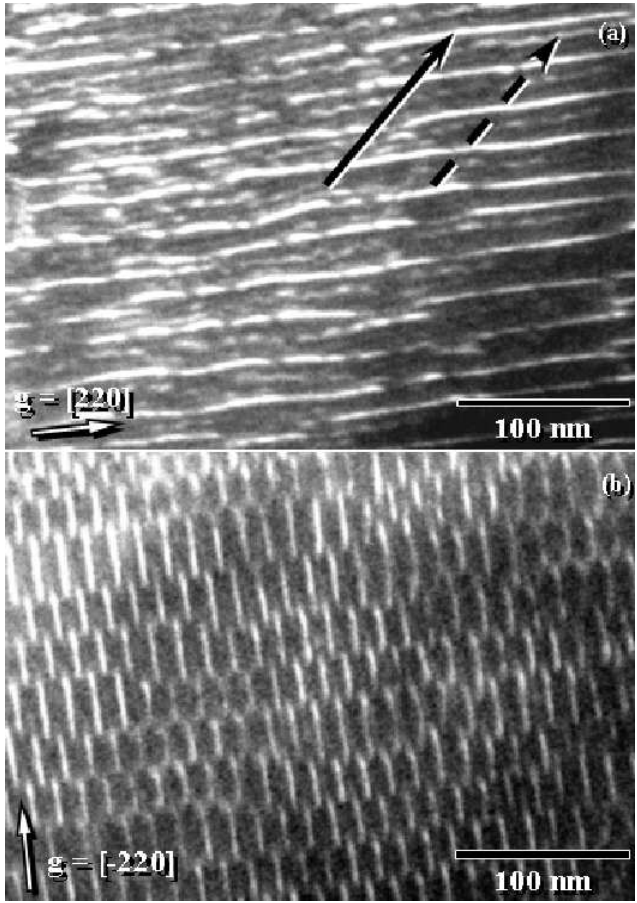

(a)

g = [220]

100 nm

(b)

g = [-220]

100 nm





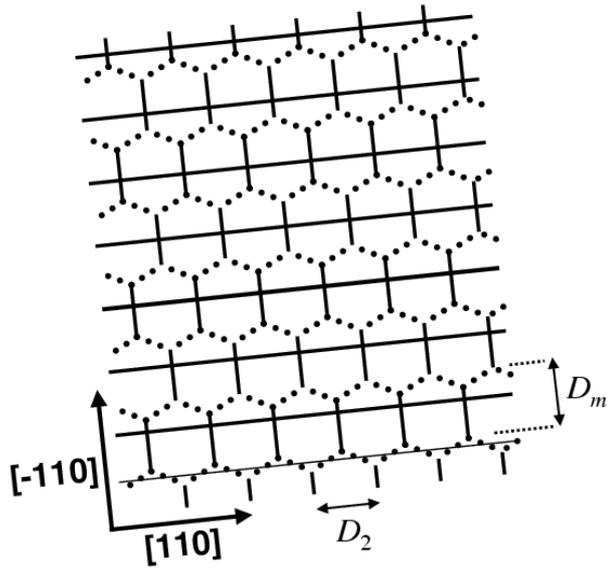



Figure 3, Coelho, Phys. Rev. B.

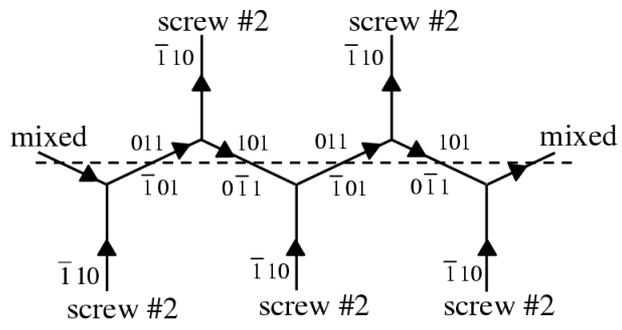

Figure 4, Coelho, Phys. Rev. B.

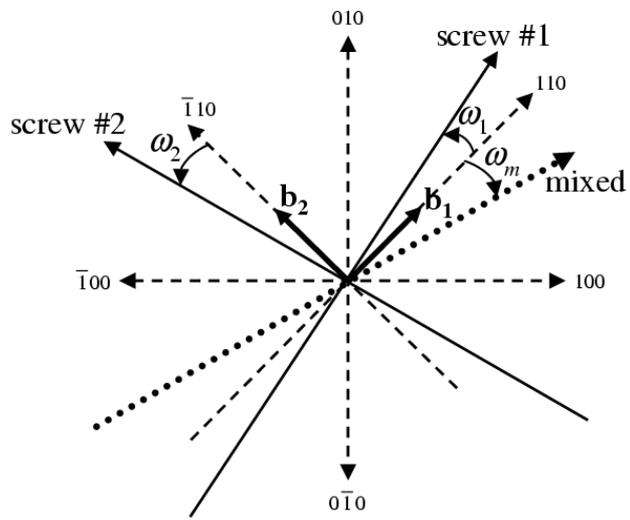



Figure 5, Coelho, Phys. Rev. B.

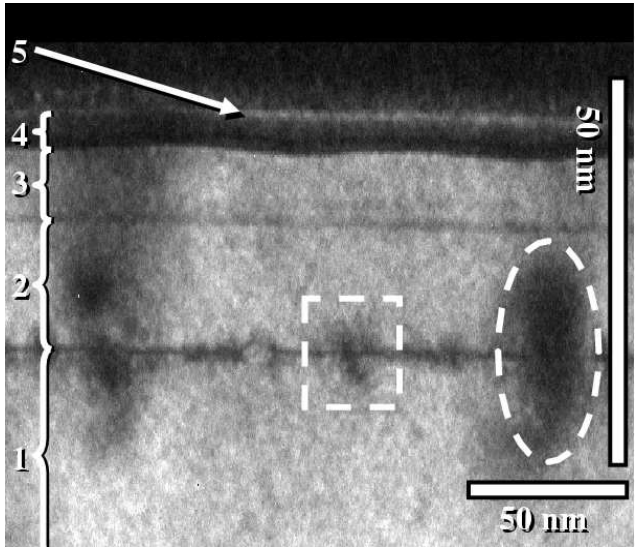



Figure 6, Coelho, Phys. Rev. B.

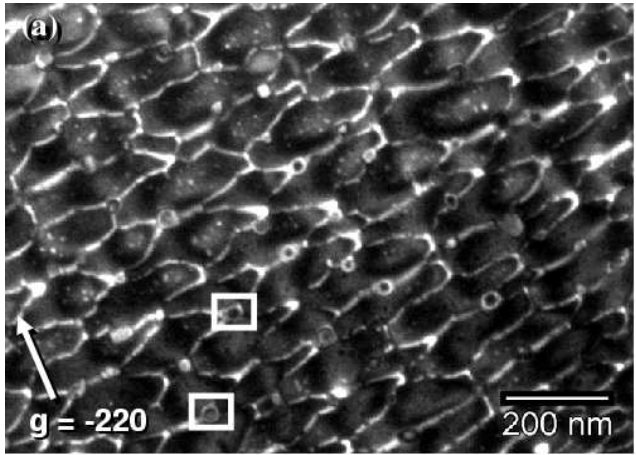

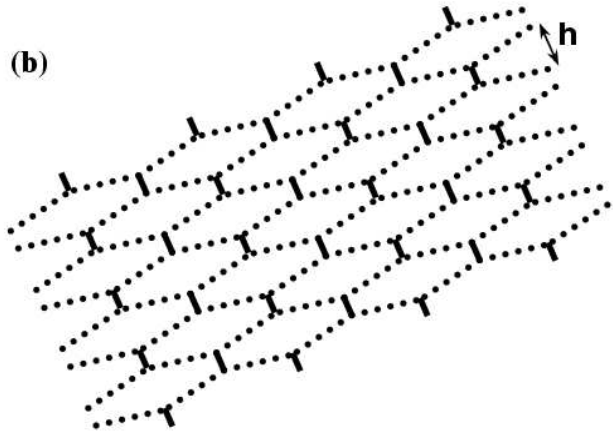



Figure 7, Coelho, Phys. Rev. B.

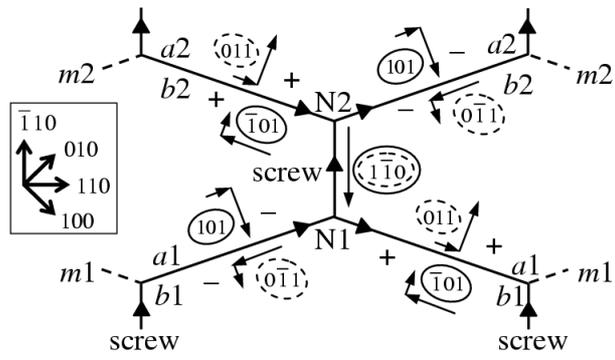



Figure 8, Coelho, Phys. Rev. B.

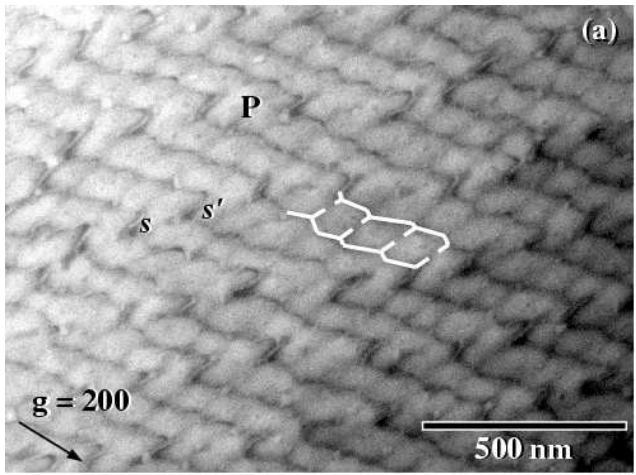

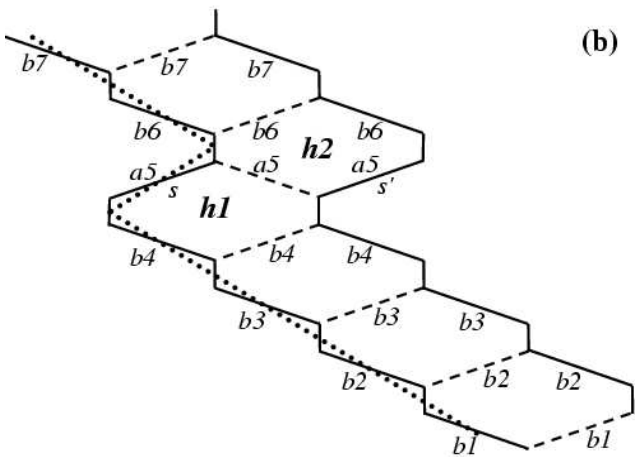



Figure 9, Coelho, Phys. Rev. B.

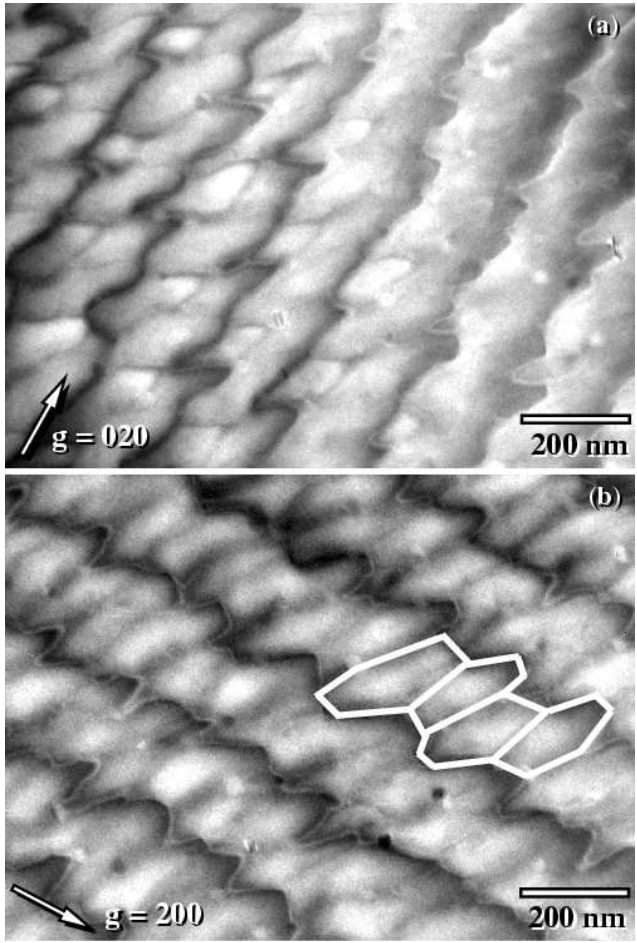



Figure 10, Coelho, Phys. Rev. B.

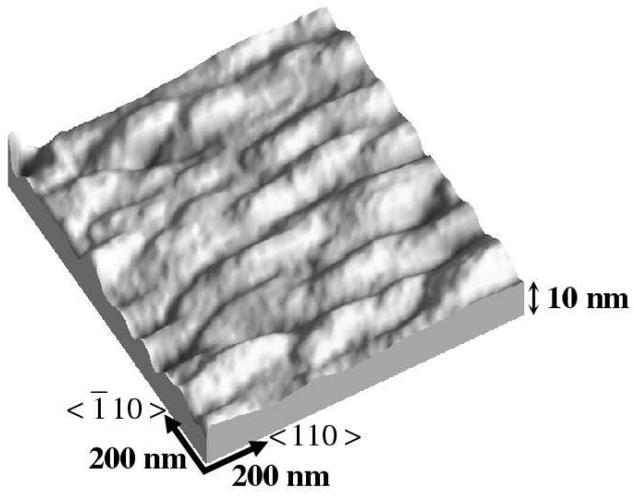



Figure 11, Coelho, Phys. Rev. B.

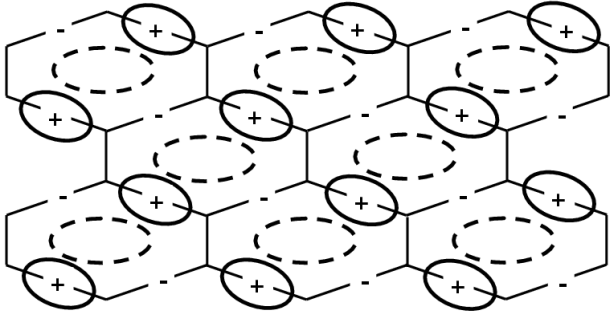